# Transient grating spectroscopy of thermal diffusivity degradation in deuterium implanted tungsten

Abdallah Reza[1*], Yevhen Zayachuk[2], Hongbing Yu[1], Felix Hofmann[1†]

[1]Department of Engineering Science, University of Oxford, Parks Road, Oxford, OX1 3PJ

[2]Department of Materials, University of Oxford, Parks Road, Oxford, OX1 3PH

## Graphical Abstract

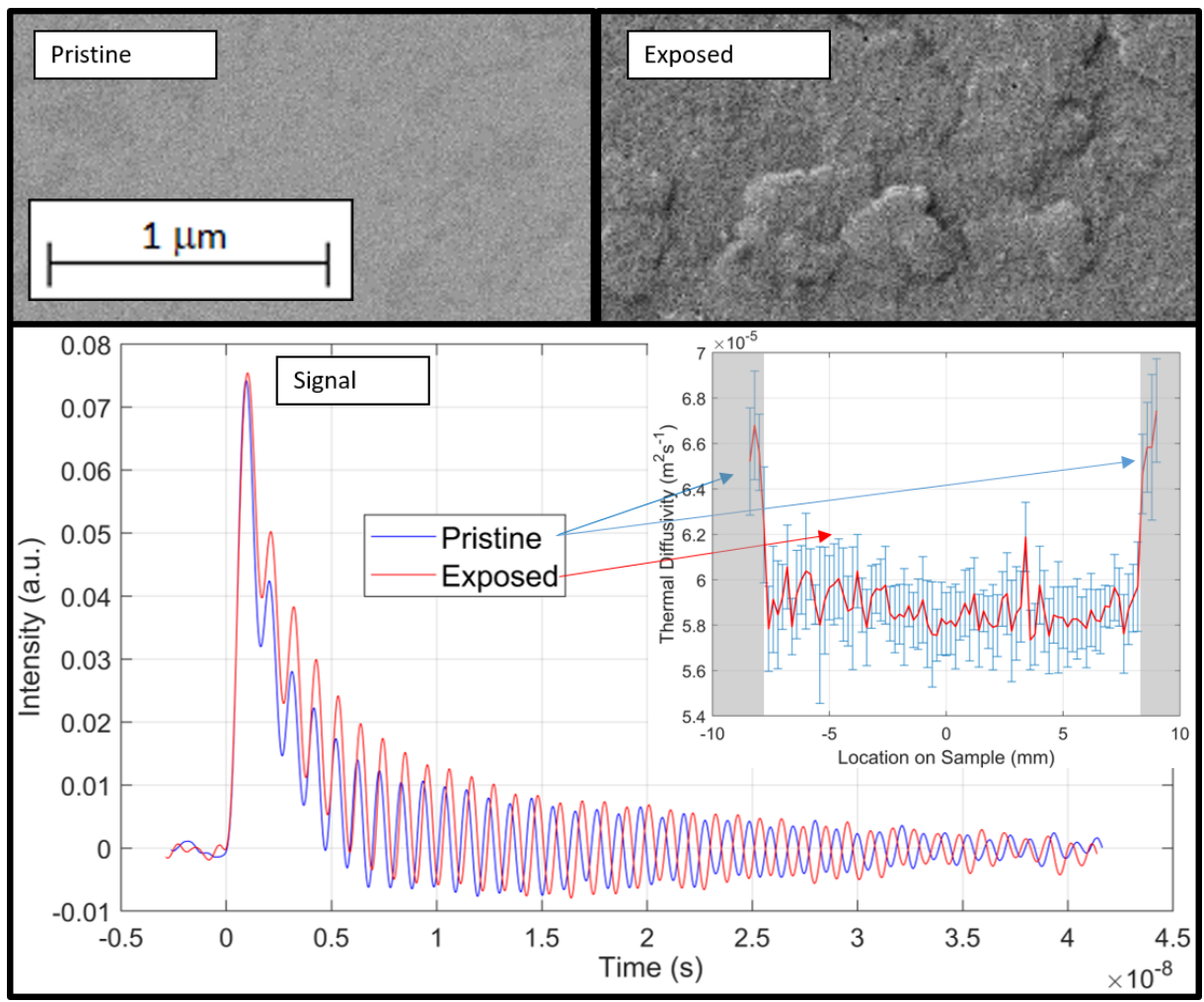

## Abstract

Using transient grating spectroscopy, we measure thermal diffusivity and surface acoustic wave speed in tungsten exposed to different fluences of deuterium plasma. Scanning electron microscopy (SEM) shows the formation of surface blisters that have similar morphology for all fluences considered. A significant reduction in thermal diffusivity and surface acoustic wave speed occurs as a result of plasma exposure. A saturation of the thermal diffusivity reduction with fluence is seen. Deuterium ion flux density appears to play a more important role in thermal diffusivity reduction than exposure time. These observations have important implications for plasma facing components in future fusion reactors.

* mohamed.reza@eng.ox.ac.uk,　†felix.hofmann@eng.ox.ac.uk

## Main Text

Mitigating climate change will require new long-term solutions for sustainable power generation. Nuclear fusion could provide a carbon-neutral and sustainable long-term solution. However, there are major engineering challenges to be addressed on the path to its commercialisation. One such challenge is the performance of plasma-facing materials, especially those used in the divertor, which will face the harshest conditions. Tungsten is the front runner for divertor armour due to its high melting point, low sputtering yield, strength at high temperature and high thermal conductivity [1]. As such, a detailed exploration of its performance under fusion reactor relevant conditions is of high importance [1–3].

Understanding material property evolution due to harsh environments is key for the success of ITER and DEMO fusion programs [3]. It is anticipated that tungsten divertor armour in a tokamak reactor will be exposed to high heat fluxes (~10 MW/m$^2$ during steady state operation and ~20 MW/m$^2$ during slow transients [4,5]), intense radiation with 14.1 MeV fusion neutrons and bombardment with ions, predominantly hydrogen and helium isotopes. A particular challenge is that hydrogen isotopes readily diffuse into the metal matrix, where they can substantially modify material properties and structure [6–9]. Trapping of deuterium at defects and self-trapping have been observed to lead to the formation of voids in the material and the emergence of surface blisters [7–9]. Since tungsten armour components will be exposed to high heat loads, this raises concerns, as blisters are expected to degrade the heat removal capability of plasma facing components by reducing near-surface thermal diffusivity. This in turn would lead to higher than predicted surface temperatures, potentially causing localised melting. In this study we quantify the effect of deuterium plasma on the thermal diffusivity of tungsten over a range of operation-relevant temperatures, flux densities and exposure times.

Deuterium injected into the material from the plasma affects the near surface region up to depths of a few microns [10]. Bulk thermal transport measurement techniques are not suitable for measuring the properties of this thin layer. Instead a different approach, called transient grating spectroscopy (TGS) [11–14] is used in this study. It allows selective probing of the properties of few-micron-thick surface layers. A detailed description of the technique is provided elsewhere [12,14,15]. Briefly, in TGS two short, coherent laser pulses (0.5 ns, 532 nm) are overlapped on the sample surface with a well-defined angle. Interferences of the beams creates an intensity grating with wavelength $\lambda$ on the sample surface. For tungsten, ~50% of the light is absorbed, with an expected penetration depth of ~9 nm [16], leading to the formation of a temperature grating in the sample. This grating decays as thermal energy diffuses from maxima to minima and into the bulk. In addition, rapid thermal expansion launches two counter-propagating surface acoustic waves (SAWs).

Both the SAWs and the temperature grating lead to a modification of the sample surface height. This "transient" grating acts as a reflective diffraction grating and can be probed by the diffraction of a second beam (continuous wave, 559.5 nm). The thermal diffusivity of a surface layer with thickness $\lambda/\pi$ determines the rate at which the transient grating decays. Hence, by analysing the amplitude decay of the diffracted probe beam, the thermal diffusivity can be determined [13].

The TGS setup used in this study incorporates the simultaneous dual heterodyne method [14]. The transient grating wavelength on the sample was $\lambda$ = 2.758 ± 0.001 µm, quantified using the SAW frequency of a tungsten calibration sample. This was chosen to get a probing depth of ~1 µm [13]. The average power at the sample was 1.5 mW for the pump (1 kHz repetition rate with 1.5 µJ per excitation) and 22 mW for the probe (88 mW optically chopped at 1 kHz with a 25% duty cycle). Given a reflectivity of ~50%, the pump and probe power absorbed were ~0.75 mW and ~11 mW respectively. The probe and excitation spots on the sample were 90 µm and 140 µm respectively ($1/e^2$ level). Measurements were performed in a vacuum of ~1 x $10^{-3}$ mbar at room temperature.

Polycrystalline tungsten discs with 20 mm diameter, 1 mm thickness and 99.95% purity were used. The sample surface was ground with SiC abrasive paper down to 2500 grit, followed by mechanical polishing with 3 and 1 µm diamond suspension on NLH polishing cloths. The samples were electrochemically polished in 0.5 wt % NaOH water solution at 12 V and heat treated at 1300 K in vacuum (~$10^{-5}$ mbar) for 1 hour. Supplementary figure S1 shows electron back scatter diffraction (EBSD) maps of the microstructure. Deuterium plasma exposures were performed at the linear plasma generator Pilot-PSI [17] (ion energy ~50 eV, Gaussian beam profile with ~10 mm FWHM). Ion flux profiles were obtained from Thomson scattering and the surface temperature was measured using an IR camera (FLIR SC7500-MB, 0.4 mm pixel size on sample). Details of the exposures can be found in [10]. TGS measurements were performed for three parallel lines across each sample with a step of 200 µm giving 80 points per line, and a spacing of 200 µm between the lines. At each position ten measurements were recorded, each averaged over 2000 pump pulses. The traces were fitted to extract thermal diffusivity and SAW speed [18]. Measurement uncertainty for the thermal diffusivity was less than 5%, and less than 0.07% for the SAW speed. Supplementary figure S2 shows a TGS trace and the fit used. Scanning electron microscopy (SEM) was performed on an Auriga Zeiss dual-beam FIB-SEM.

To investigate how thermal diffusivity varies with deuterium flux, fluence and exposure temperature, an experimental matrix with four exposure conditions was used: high temperature high dose (HT HD), high temperature low dose (HT LD), low temperature low dose (LT LD) and low temperature high dose (LT HD). "LT" refers to ~450 K, "HT" to ~650 K, "LD" to ~5 x $10^{25}$ $m^{-2}$ (70 seconds exposure), and "HD" to ~$10^{27}$ $m^{-2}$ (1400 seconds exposure). The Gaussian nature of the ion beam allowed a spatial variation of flux and fluence across the sample. A 2 mm wide region around the sample edge was shielded from the plasma by a molybdenum clamping ring. Fig. 1 shows the deuterium ion fluence profiles and temperature profiles for all samples.

During the exposure, the central region of the two samples exposed at high temperature was contaminated with molybdenum originating from the plasma source. This contamination was characterised using energy-dispersive x-ray spectroscopy (EDX) (see supplementary section on impurity characterisation). CASINO [19] was used to estimate the electron probing depths for different energies. The centre of the HT HD sample, showed ~60% molybdenum for a 280 nm probing depth. The HT LD sample showed ~6% molybdenum for the same depth. Away from the centre and in the unimplanted regions, both samples showed no molybdenum even for a 39 nm probing depth. The LT HD sample had 2% molybdenum at the centre with the same probing depth and insignificant amounts at further in. Hence, given the TGS probing depth of ~1 µm, the molybdenum surface contamination is expected to have little effect except in the central regions of the HT samples, which have been excluded from this study (boxed region in Fig. 1).

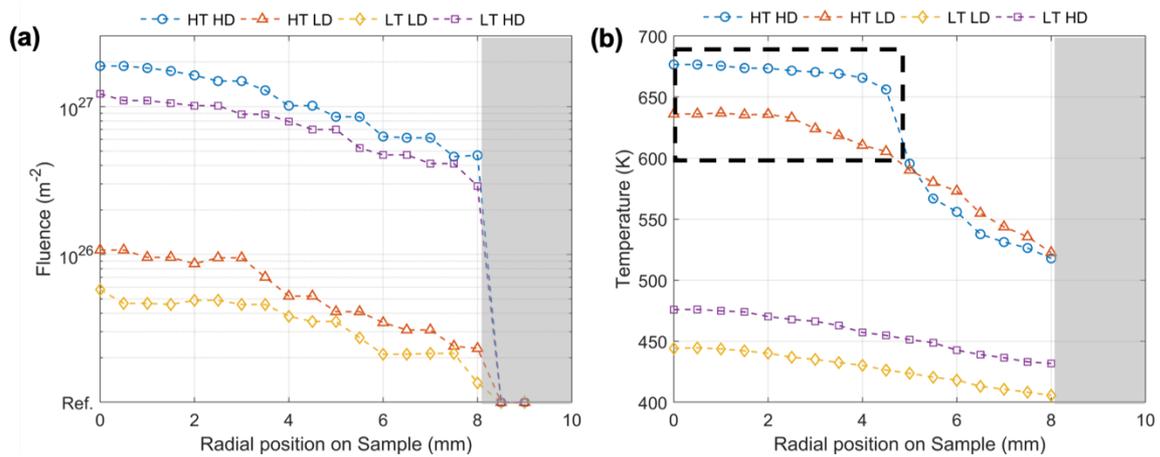

**Figure 1.** (a) Profile of incident deuterium ion fluence for the four samples. The reference fluence at the shielded sample edges is zero. (b) Surface temperature profiles for the four different samples. The temperature of the shielded edges is assumed to be consistent with the near-Gaussian nature of the temperature profile. Boxed region in the centre depicts the molybdenum-contaminated regions. The shielded region is depicted by grey shading.

SEM micrographs of the exposed regions show prominent blistering in all samples (Fig. 2), while unexposed regions are blister free (see supplementary Fig. S3). For the LT samples there is no obvious spatial variation of blister morphology. Blister diameters are ~0.5 µm, consistent with previous studies [20]. Micrographs from the LT-LD sample (Fig. 2 (a) and (b)) and the LT-HD sample (Fig. 2 (c) and (d)) shows similar blistering, despite the fact that the LT-HD sample received a dose 20 times higher than the LT-LD sample. This suggests that, at least for the conditions considered here, the blistering effects is largely independent of deuterium flux and fluence. The micrographs of samples exposed to deuterium at the higher temperature (HT) (Fig. 2 (e) and (f)) show blisters that are somewhat smaller. However, the morphology seems to vary little with deuterium dose. Cross-sectional micrographs of the LT LD sample showed that blisters extend up depths of 0.8 – 1 µm (See supplementary figures S4-S5).

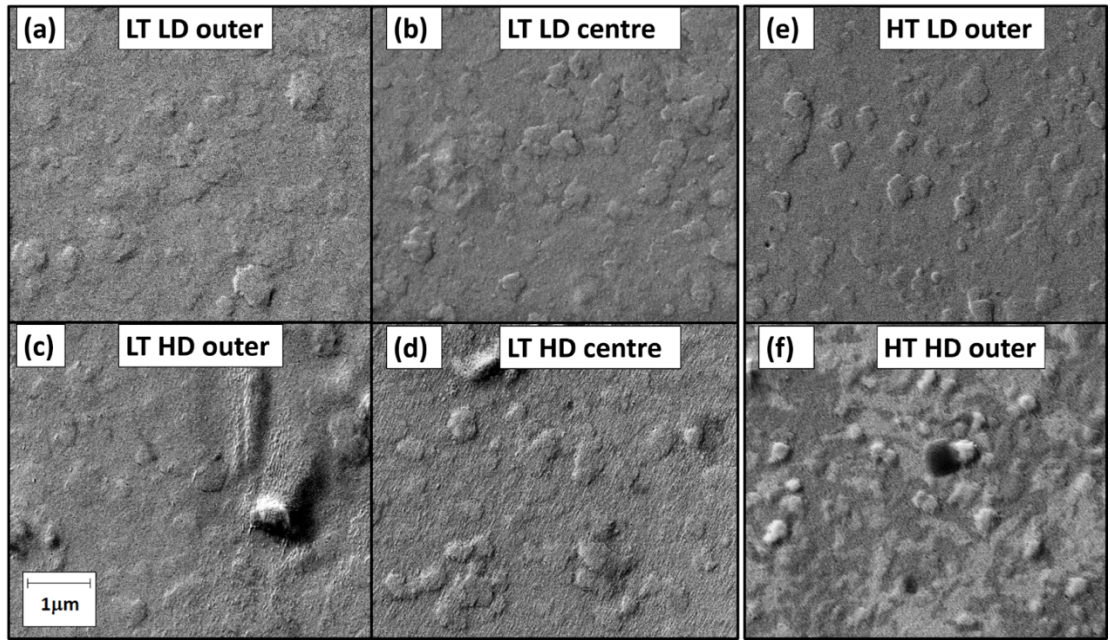

**Figure 2.** SEM micrographs of the exposed samples at different radial locations 'r'. (a) LT LD r = 5mm, (b) LT LD r = 0 mm, (c) LT HD r = 5 mm, (d) LT HD r = 0 mm, (e) HT LD r = 5 mm, (f) HT HD r = 5 mm. The same scale bar applies to all micrographs.

The TGS scans included the unexposed sample edges as a built-in reference. Fig. 3 shows thermal diffusivity and SAW speed as a function of position in the LT-HD sample. Profiles for the other samples are shown in supplementary figures S6-S8. The unexposed edges are identified by their higher thermal diffusivity and SAW speed (Fig. 3). The changes between unexposed and exposed regions are also evident from TGS traces representative of both areas (supplementary Fig. S9). The unexposed regions (Fig. 3) have a thermal diffusivity of (6.6 ± 0.2) x $10^{-5}$ $m^2s^{-1}$, which agrees well with literature values for pristine tungsten [21–24]. Thermal diffusivity in the exposed region reduces to between 5.5 and 6.1 x $10^{-5}$ $m^2s^{-1}$.

At the centre of the LT-HD sample there is a sharp drop in thermal diffusivity to 5.0 x $10^{-5}$ $m^2s^{-1}$. Since the samples were exposed to temperature variations and deuterium beams with near-Gaussian shape, this feature is surprising. This drop is the consequence of nuclear reaction analysis (NRA) previously performed on these samples [10]. The NRA study involved irradiation with helium ions to a damage level of 3.2 dpa and is seen to have a substantial effect on the thermal diffusivity. These points were hence excluded from the present analysis of deuterium-exposure effects.

The SAW speed of the unexposed edges is 2668 ± 2 $ms^{-1}$, in good agreement with previous measurements [25]. The SAW speed drops steeply from the unexposed to exposed region and then continues to gradually decrease towards the centre of the sample. The oscillations of SAW speed with a period of ~0.8 mm in the exposed region are reliably found in repeated measurements. Such an oscillation is not evident in either the flux or temperature profiles (Fig. 1). Therefore, it cannot be the result of an oscillation in flux or temperature. It is also not present in the thermal diffusivity results, hence it is not a measurement error. Thus, it is likely that this is an effect of the grain dependent elastic anisotropy. This variation has a magnitude of ~10 $ms^{-1}$, which agrees with previous work [25]. EBSD scans (see supplementary figure S1) show grain sizes of 200 -400 μm which is comparable to the period of these oscillations, reaffirming this.

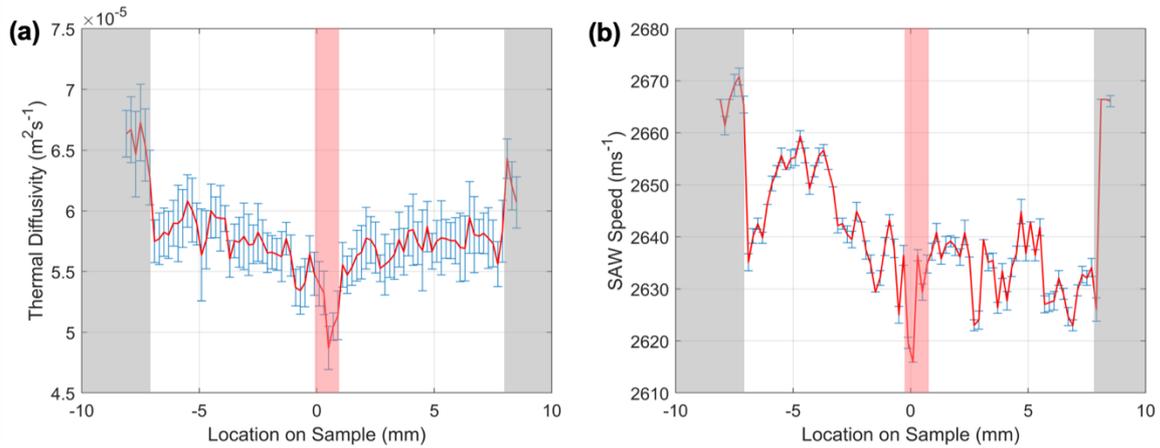

**Figure 3.** Thermal diffusivity (a) and SAW frequency (b) profiles for the low temperature low dose (LT HD) sample. Error bars are the standard deviation for each point, obtained over 10 measurements. The x-axis is corrected for the centre of the incident ion beam. The shielded region is depicted by the grey box. Regions excluded from this study due to previous NRA measurements are shaded in red.

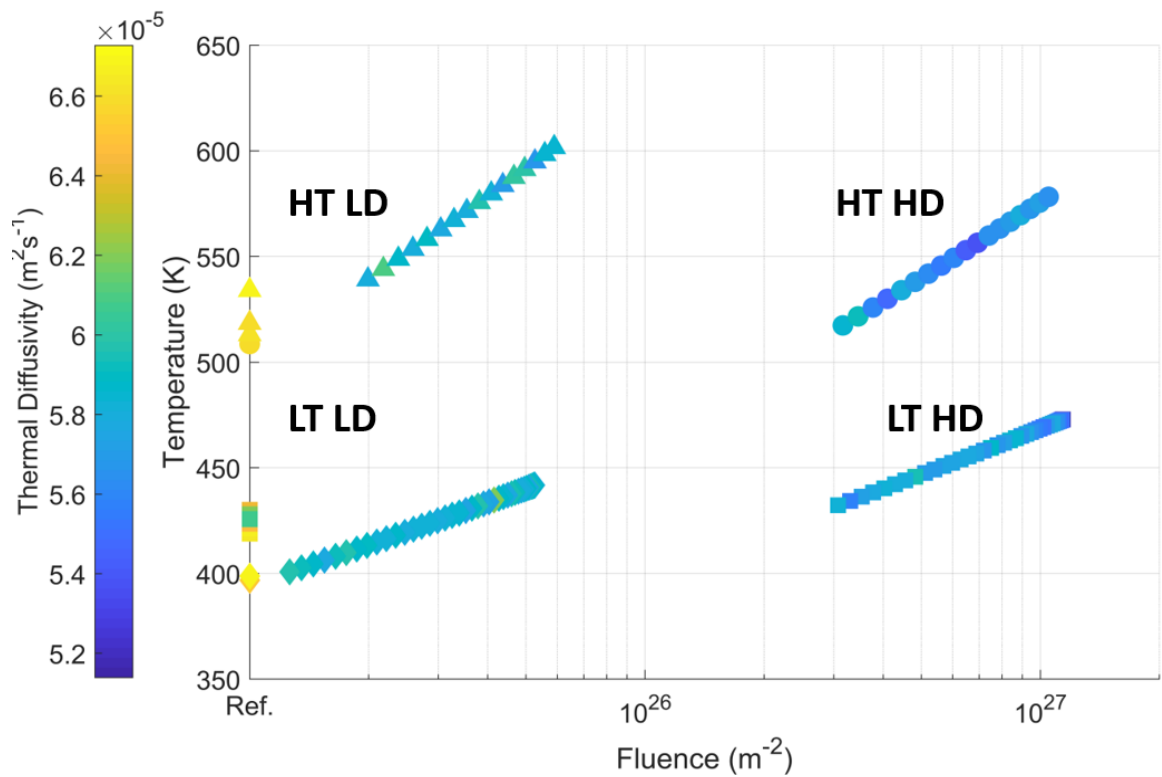

**Figure 4.** Measured thermal diffusivity with fluence and temperature for the four samples considered. Data from the high temperature high dose sample (HT HD) is depicted by circles, high temperature low dose (HT LD) by triangles, low temperature low dose (LT LD) by diamonds and low temperature high dose (LT HD) by squares. Reference fluence is zero, which is the value for the unexposed region. An alternative representation of this figure, with error bars, is provided as supplementary figure S10.

Fig. 4 summarises the thermal diffusivity measurements, plotted as a function of exposure temperature and fluence. The unimplanted regions in all four samples have thermal diffusivity greater than 6.5 x $10^{-5}$ m$^2$s$^{-1}$ in agreement with previous studies [21–24]. There is a significant drop in thermal diffusivity between virgin and deuterium-exposed areas, irrespective of fluence and exposure temperature. The high dose (HD) samples show a larger reduction in thermal diffusivity compared to their low dose counterparts. The drop in diffusivity between the LD and HD samples (from $6.0 \times 10^{-5}$ $m^2s^{-1}$ to $5.5 \times 10^{-5}$ $m^2s^{-1}$), is similar in magnitude to the drop between the LD samples and unexposed regions (from $6.6 \times 10^{-5}$ $m^2s^{-1}$ to $6.0 \times 10^{-5}$ $m^2s^{-1}$). However, HD samples received a 20 times higher dose than LD samples, meaning that the drop in thermal diffusivity is not proportional to fluence.

A similar trend has been observed by NRA, where the sub-surface deuterium concentration in deuterium plasma-exposed tungsten samples was measured for different fluences [10]. NRA profiles showed a large difference in deuterium concentration between unimplanted regions and regions exposed to a fluence of ~$5 \times 10^{25}$ $m^{-2}$, similar to the exposures considered here. The deuterium concentration in the samples exposed to ~$10^{27}$ $m^{-2}$ was only slightly higher. This suggests that with increasing fluence the deuterium concentration, blistering and thermal diffusivity reduction saturate. The fluence at which this saturation occurs may vary and depends on a number of factors such as the crystallography, heat treatment and surface polishing [7,8]. Since we observe little change in thermal diffusivity when fluence is increased by more than one order of magnitude, it is likely that our fluences are within this saturation regime. The thermal diffusivity seems to correlate quite well with blister morphology, which remains largely unchanged between low and high dose exposed samples. Previous studies [26] reported an increase in blistering with fluence for fluences of ~$10^{23}$ -$10^{25}$ $m^{-2}$. This suggests that the initial increase in blistering with fluence is followed by the saturation observed in this study. In the HT HD sample, with increasing temperature and fluence there is an initial decrease in thermal diffusivity followed by a recovery, which is not evident in the LT HD sample. Previous studies [27] showed decreased near-surface (< 1 μm) deuterium concentrations at temperatures above 490 K. Hence this feature is believed to be an effect of the elevated temperature, which has initiated a recovery of thermal diffusivity, despite the higher D fluence.

Previous studies suggest different mechanisms for thermal diffusivity reduction due to plasma exposure/ion implantation [18,23,24,28,29]. Defects created by high energy ions or neutrons can function as effective electron scattering sites, thereby reducing thermal diffusivity, which for tungsten at room temperature is dominated by electron-mediated transport. Significant reductions in thermal diffusivity have been reported for self-ion implanted tungsten [28,29]. However, the deuterium ion energy used in this study (~50 eV) was much lower than usual self-ion implantation energies (150 keV - 30 MeV)[29,30], and below the threshold required to produce displacement damage [31]. Hence displacement damage does not play a prominent role. However studies have shown that foreign gaseous ions such as deuterium and helium infused into the tungsten matrix can self-trap, i.e. an interstitial deuterium ion is attracted to another [18,32–35]. These ions then cluster, forming bubbles with increasing pressure, resulting in blisters. From the SEM micrographs in figure 2, it is clear that there is extensive surface blistering in all samples. These blisters are surface manifestations of such sub-surface gas-filled cavities [36]. These act as additional scattering centres and also suppress thermal diffusivity by reducing thermal contact between the surface and the bulk

Our results have important implications for the design of armour components in future fusion reactors. In test components pristine tungsten was observed to undergo extensive melting at heat fluxes of 27-30 $MW/m^2$ [4], while future fusion reactors are expected to have peak heat fluxes of 10-20 $MW/m^2$ in the divertor [5]. A reduction of thermal diffusivity will decrease the maximum heat flux that can be accommodated without melting. As such it is key that degradation of thermal transport properties is accounted for in the design of armour components. A positive conclusion from our results is that the deuterium-exposure-induced reduction in thermal diffusivity appears to saturate quickly, rather than continuing to progress with extended exposure.

All raw and processed data, SEM images, and TGS processing scripts used in this study are available online (https://doi.org/10.5281/zenodo.3362429).

## Acknowledgements

The authors are grateful to Cody A. Dennett and Michael P. Short for help with the implementation of the TGS setup. We acknowledge funding from the European Research Council (ERC) under the European Union's Horizon 2020 research and innovation programme (grant agreement No. 714697). Plasma exposures were performed within the framework of the EUROfusion Consortium and received funding from the Euratom research and training programme 2014-2018 under grant agreement No 633053. The views and opinions expressed herein do not necessarily reflect those of the European Commission.

# Transient grating spectroscopy of thermal diffusivity degradation in deuterium implanted tungsten: Supplementary Information

Abdallah Reza[1*], Yevhen Zayachuk[2], Hongbing Yu[1], Felix Hofmann[1†]

[1]Department of Engineering Science, University of Oxford, Parks Road, Oxford, OX1 3PJ

[2]Department of Materials, University of Oxford, Parks Road, Oxford, OX1 3PH

Supplementary Figures

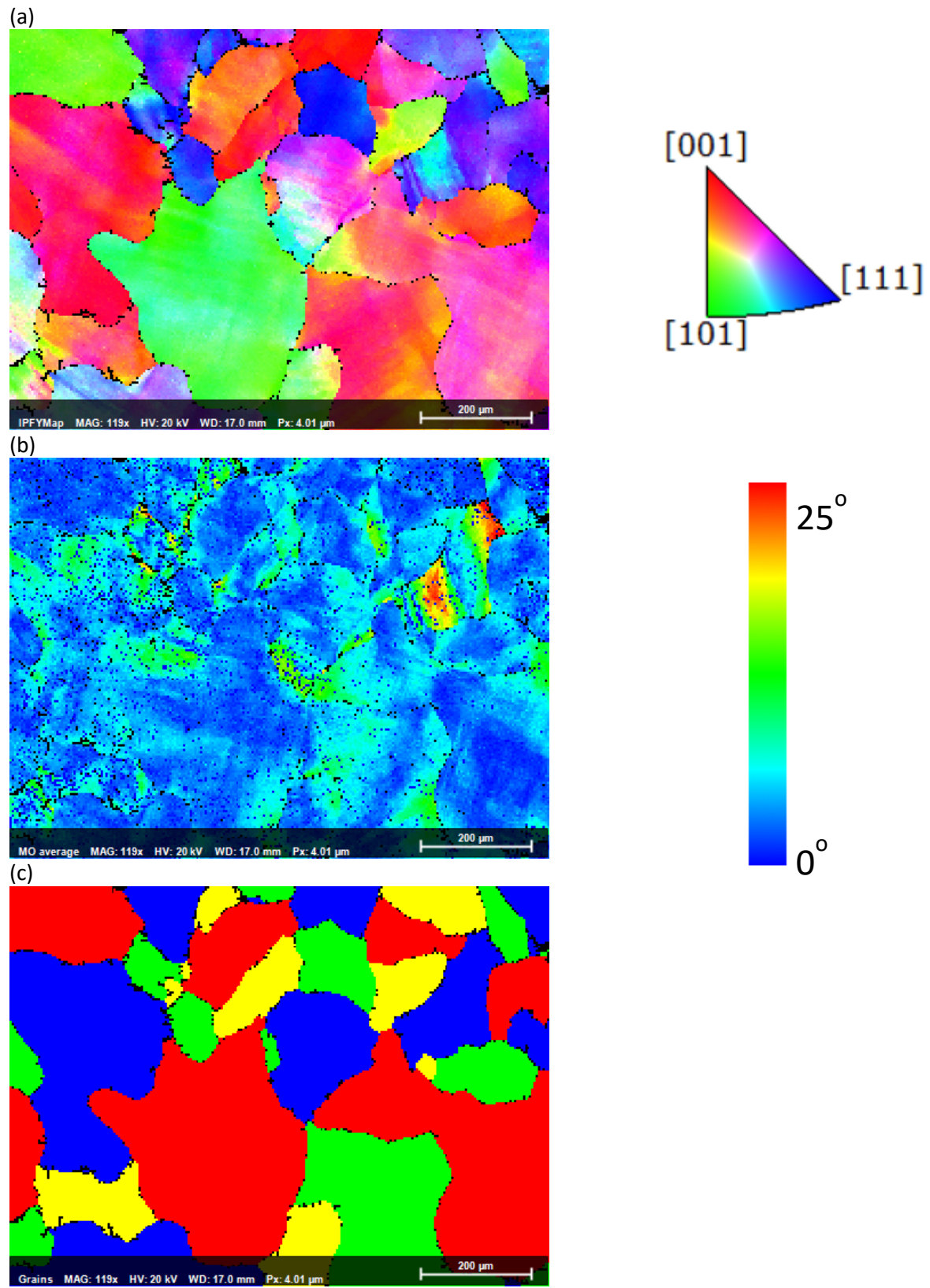

Supplementary Figure S1: Microstructure of the investigated tungsten following heat treatment, determined by EBSD: (a) Orientation map (IPFY); (b) misorientation map; (c) identified grains.

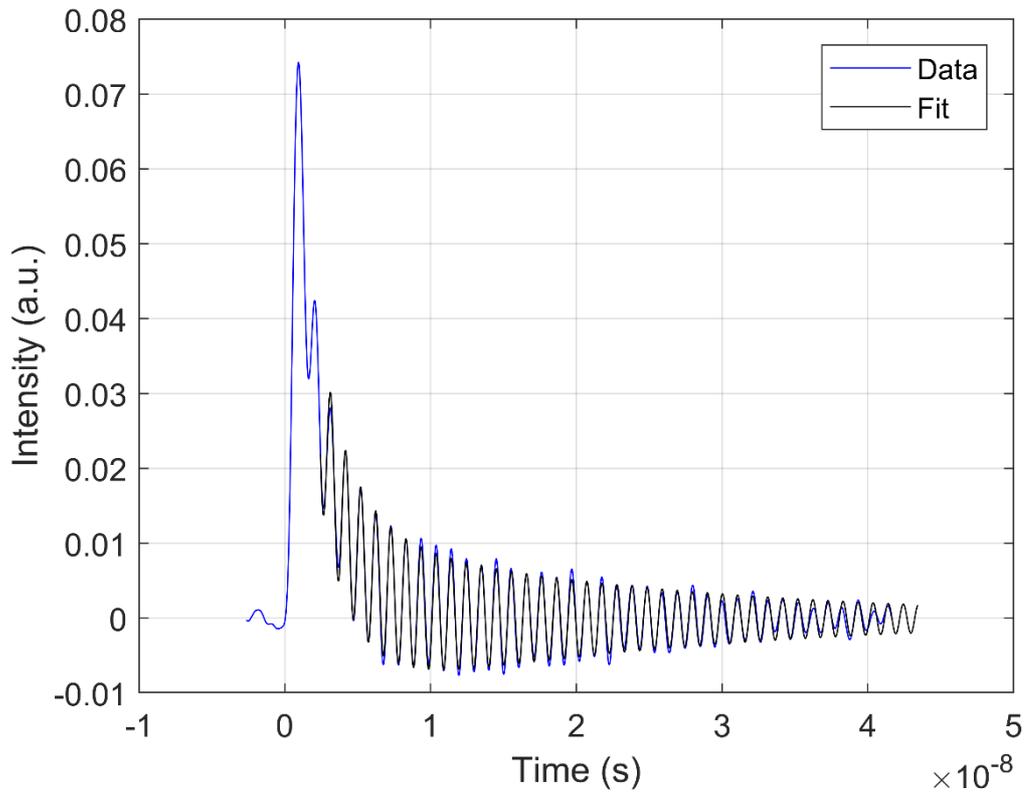

Supplementary Figure S2: Transient grating signal and fitted trace for the unimplanted region in the low temperature high dose (LT HD) sample

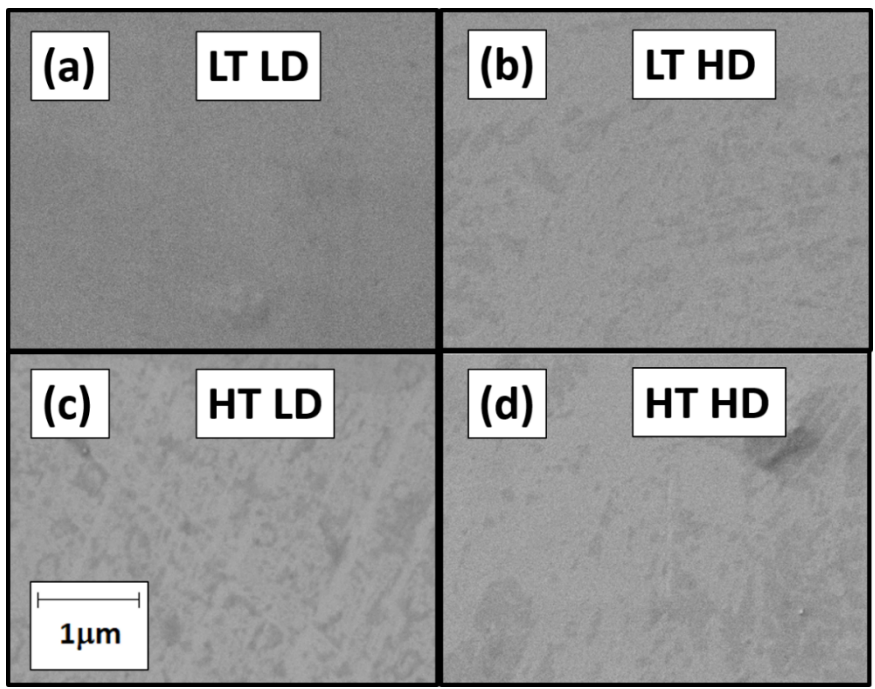

Supplementary Figure S3: SEM micrographs for the unimplanted regions in the LT LD (a), LT HD (b), HT LD (c) and HT HD (d) samples.

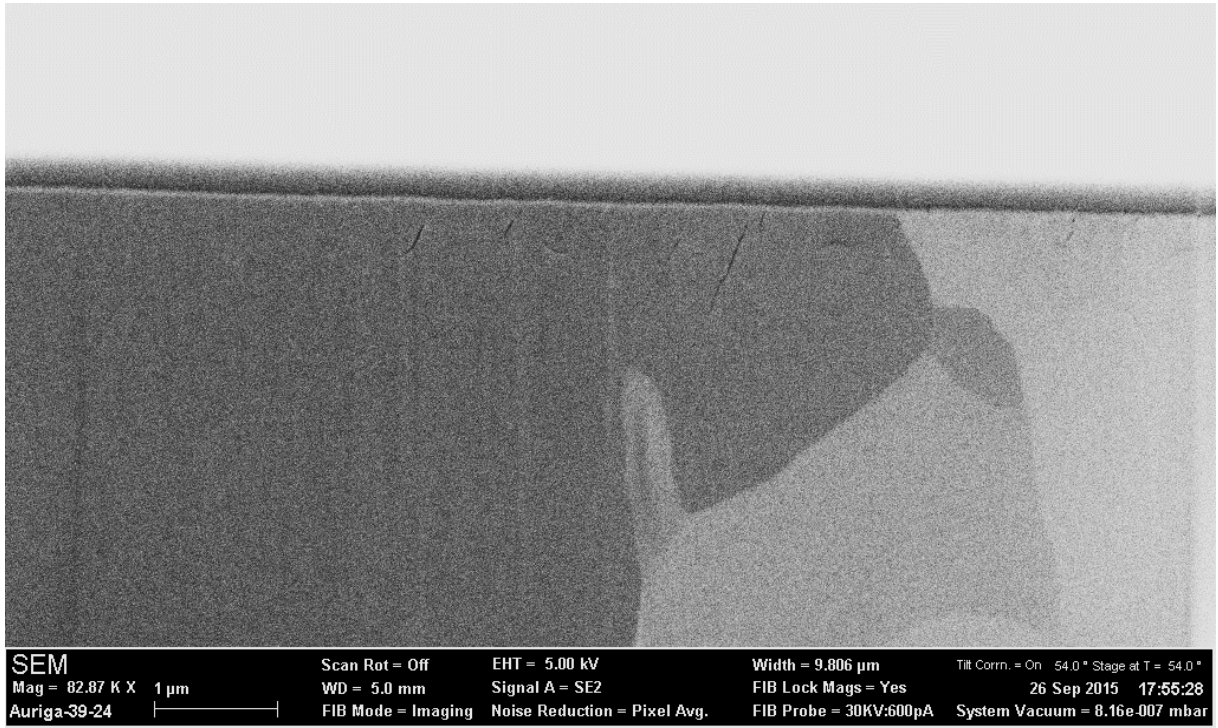

Supplementary Figure S4: Cross-sectional micrograph of the low temperature low dose (LT LD) sample at 82.87K zoom. The presence of sub –surface cavities is evident up to depths of ~1 µm.

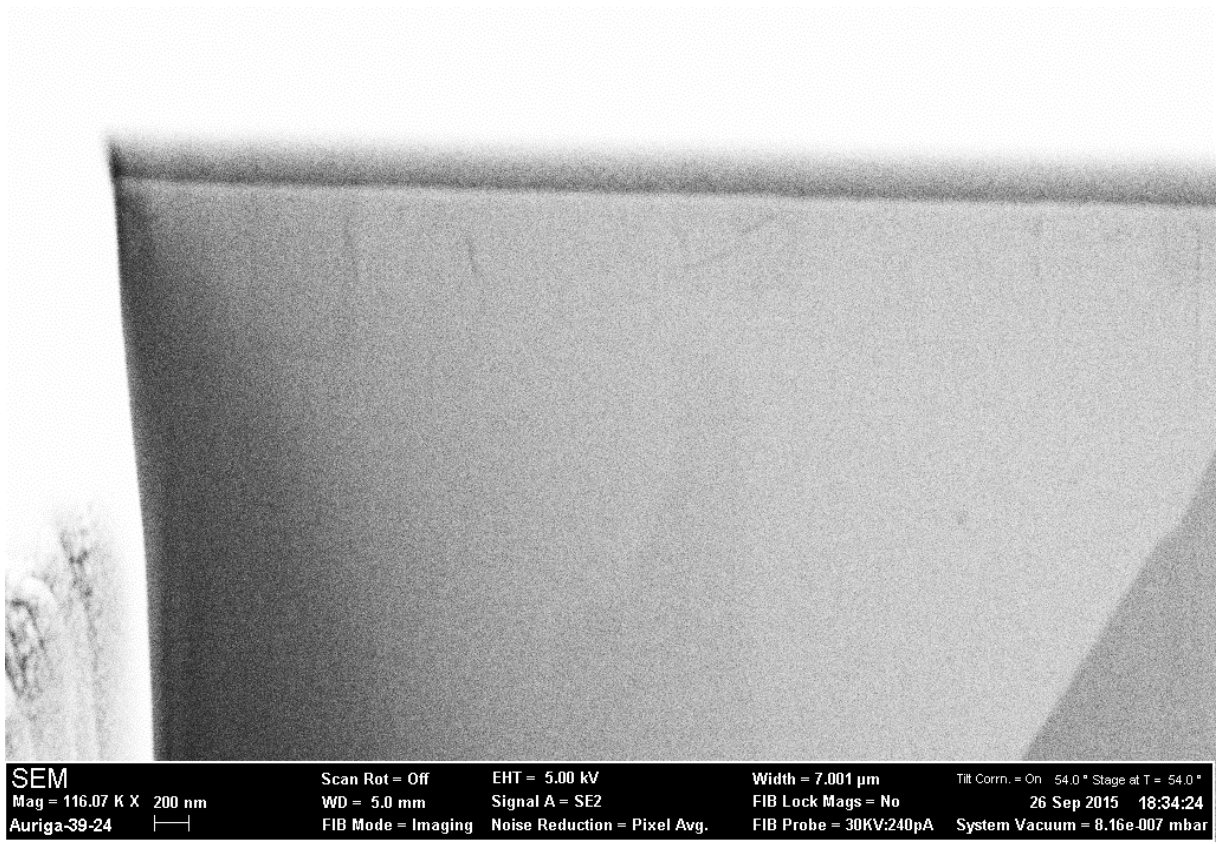

Supplementary Figure S5: Cross-sectional micrograph of the low temperature low dose (LT LD) sample at 116K zoom. The presence of sub–surface cavities is evident up to depths of ~ 1 µm.

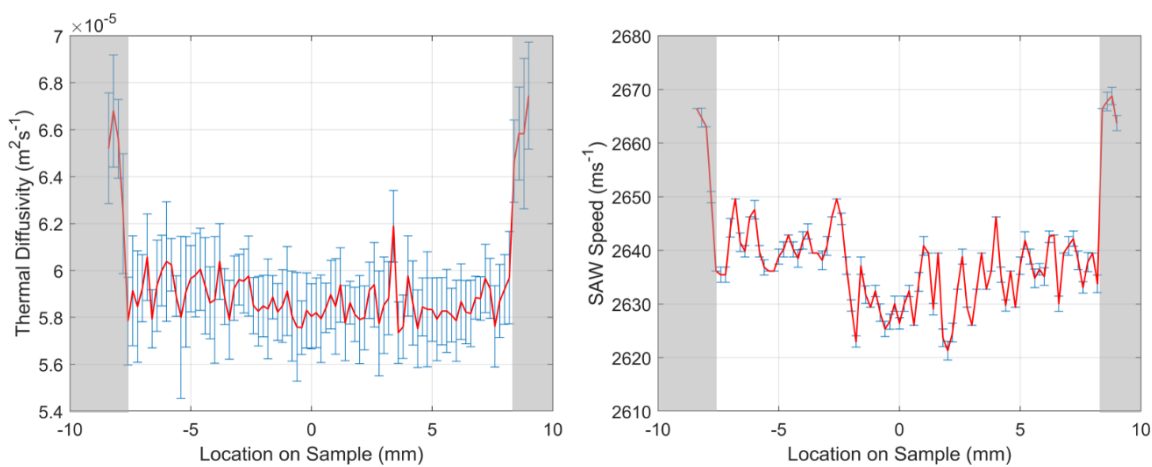

Supplementary Figure S6: Thermal diffusivity and SAW profile for the low temperature low dose (LT LD) sample. Error bars are the standard deviation for each point, obtained over 10 measurements. The red shaded region depicts data excluded from the analysis due to surface contamination. The grey shaded region was not exposed to deuterium.

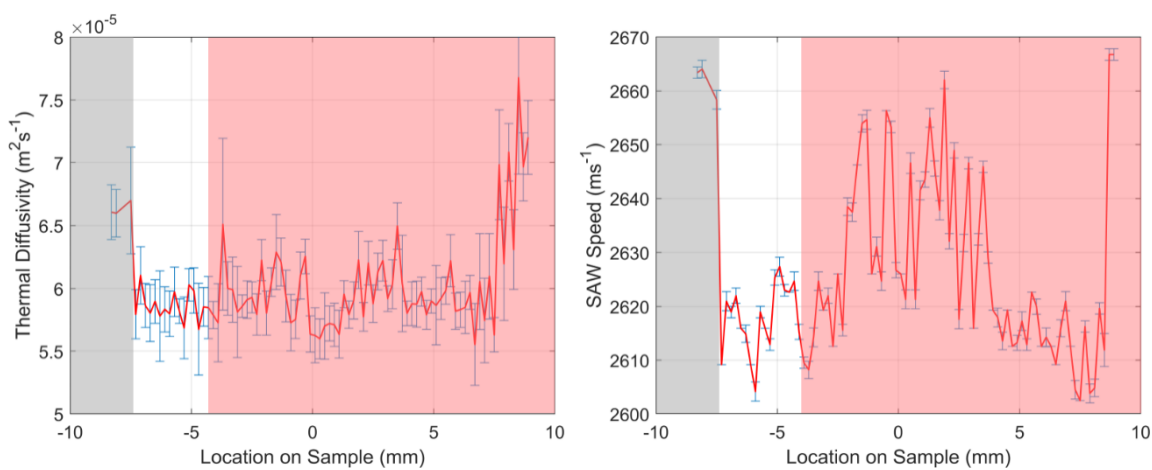

Supplementary Figure S7: Thermal diffusivity and SAW profile for the high temperature low dose (HT LD) sample. Error bars are the standard deviation for each point, obtained over 10 measurements. The red shaded region depicts data excluded from the analysis due to surface and molybdenum contamination. The grey shaded region was not exposed to deuterium.

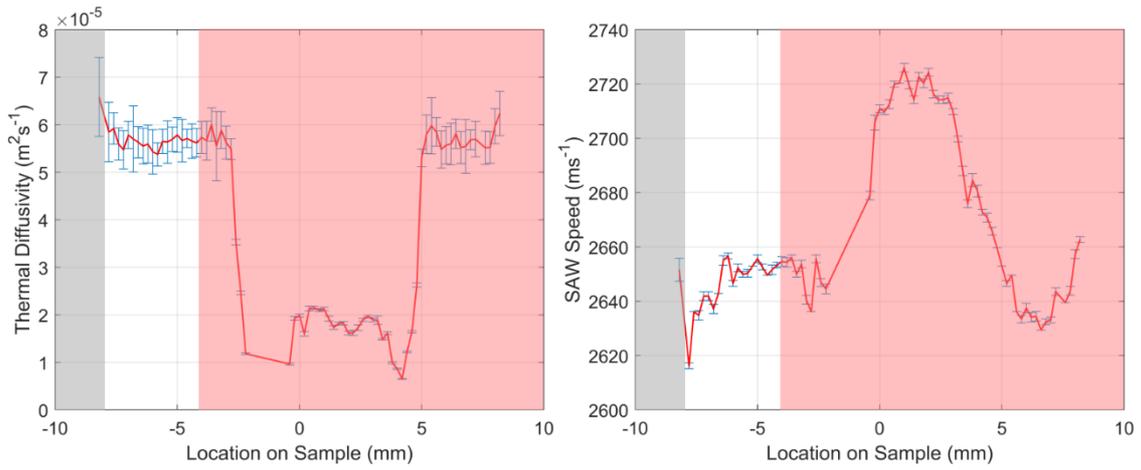

Supplementary Figure S8: Thermal diffusivity and SAW profile for the high temperature high dose (HT HD) sample. Error bars are the standard deviation for each point, obtained over 10 measurements. The red shaded region depicts data excluded from the analysis due to molybdenum contamination. The grey shaded region was not exposed to deuterium.

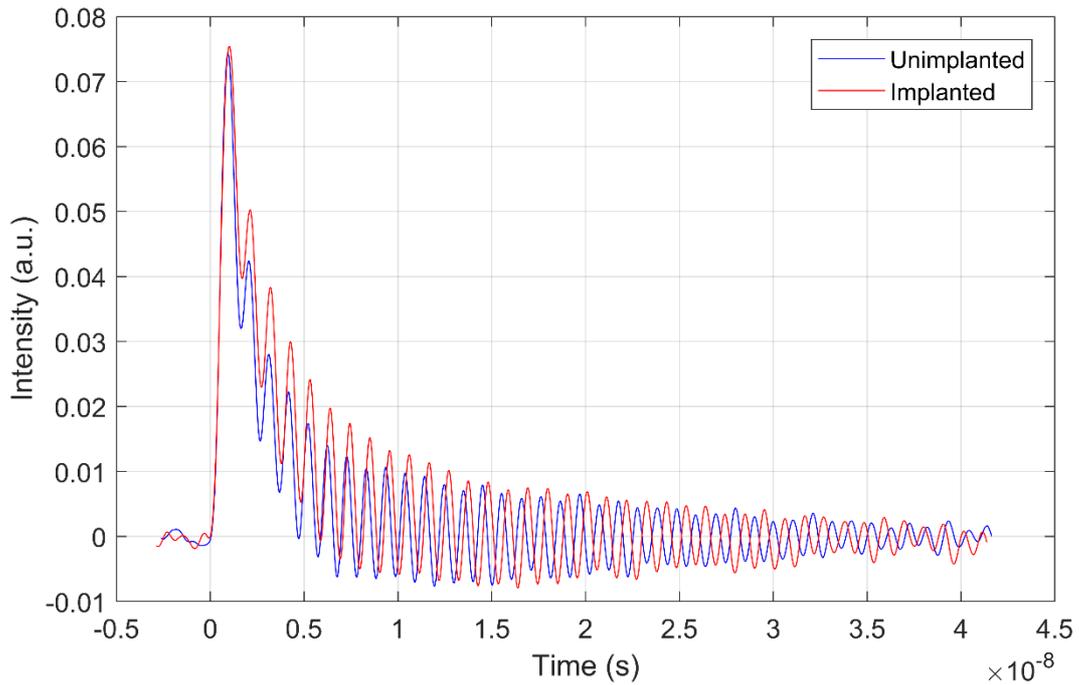

Supplementary Figure S9: Transient grating signal for a single measurement at the unimplanted edge (blue) and implanted centre (red) of the LT HD sample

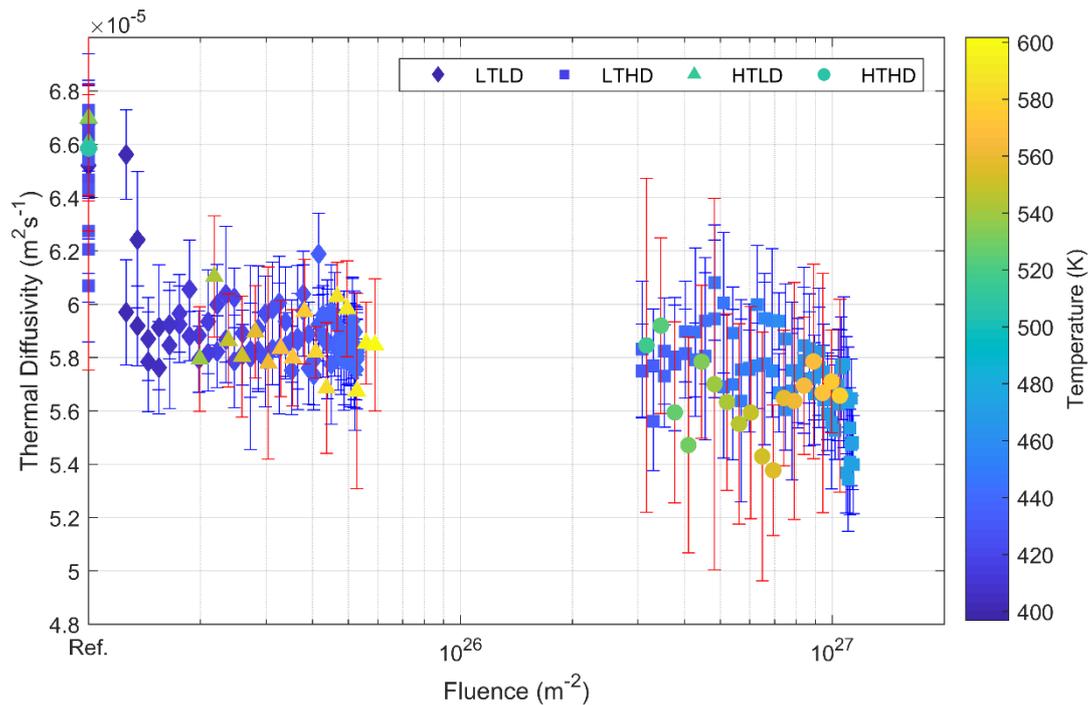

Supplementary Figure S10: Measured thermal diffusivity with fluence and temperature for the four samples considered. Data from the high temperature high dose sample (HT HD) is depicted by circles, high temperature low dose (HT LD) by triangles, low temperature low dose (LT LD) by diamonds and low temperature high dose (LT HD) by squares. Reference fluence is zero, which is the value for the unexposed region. Error bars given are the standard deviation for 20,000 measurements.

## Impurity Characterisation

Supplementary Table 1. Tungsten and molybdenum concentrations obtained by energy-dispersive x-ray spectroscopy with varied probing depth. Probing depths were estimated using simulations in CASINO [1]. The probing depths given for the 15 and 20 keV scenarios are the average of the probing depths for the 2 characteristic x-rays, MV and LIII.

| Electron energy (keV) | Probing Depth ($1/e^2$ nm) | Centre | | 5 mm off-centre | | Unimplanted | |
|---|---|---|---|---|---|---|---|
| | | W (at%) | Mo (at%) | W (at%) | Mo (at%) | W (at%) | Mo (at%) |
| HT HD | | | | | | | |
| 5 | 39 | 11.1 | 88.9 | 99.8 | 0.2 | 100.0 | 0.0 |
| 10 | 112 | 17.7 | 82.3 | 100.0 | 0.0 | | |
| 15 | 169 | 31.4 | 68.6 | | | | |
| 20 | 280 | 39.8 | 60.2 | | | | |
| HT LD | | | | | | | |
| 5 | 39 | 60.7 | 39.3 | 99.8 | 0.2 | 100.0 | 0.0 |
| 10 | 112 | 86.0 | 14.0 | | | | |
| 15 | 169 | 91.6 | 8.4 | | | | |
| 20 | 280 | 94.2 | 5.8 | | | | |
| LT HD | | | | | | | |
| 5 | 39 | 97.9 | 2.1 | | | | |
| 10 | 112 | 99.5 | 0.5 | | | | |
| 15 | 169 | | | | | | |
| 20 | 280 | | | | | | |

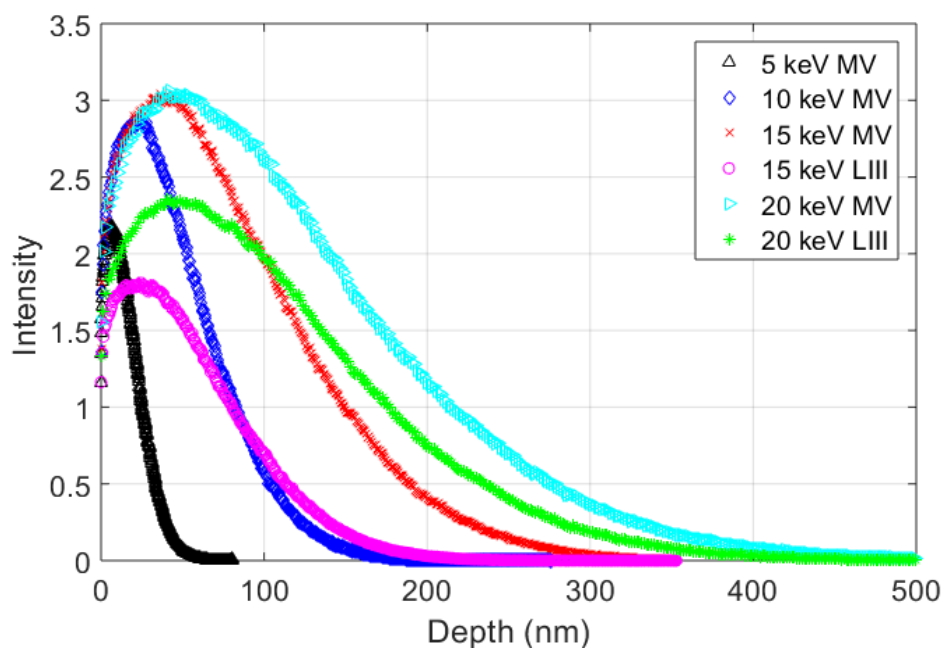

Supplementary Figure S10. EDX X-ray generation depths for the different electron energies considered. MV and LIII denote the two characteristic x-rays generated for tungsten.

## Supplementary References